\documentclass{article} 
\usepackage{geometry}

\usepackage{geometry}
\usepackage[export]{adjustbox}
 \usepackage{amsfonts}
\usepackage{amsmath}
\usepackage{authblk}
\usepackage{booktabs}
\usepackage{graphicx}
\usepackage{url}
\usepackage[margin=0.28in]{caption}
\usepackage{natbib}
\usepackage{multirow}
 \newcounter{thm}
 \newcounter{ex}
 \newcounter{re}
\providecommand{\keywords}[1]{\textbf{\textit{Index terms---}} #1}

\author[1]{Erfan Sayyari \thanks{esayyari@ucsd.edu}}
\author[2]{James B. Whitfield \thanks{jwhitfie@life.illinois.edu}}
\author[1]{Siavash Mirarab\thanks{smirarab@ucsd.edu}}
\affil[1]{Department of Electrical and Computer Engineering, University of California at San Diego, 9500 Gilman Drive, La Jolla, CA 92093.}
\affil[2]{Department of Entomology,
University of Illinois at Urbana-Champaign, 505 S. Goodwin Avenue, Urbana, IL 61801}

\title{DiscoVista: interpretable
visualizations of gene tree discordance \footnote{Accepted by Molecular Phylogenetics and Evolution, 2018}}
\begin{document}
\maketitle






\begin{abstract}
Phylogenomics has ushered in an age of discordance. 
Analyses often reveal abundant discordances among phylogenies of different parts of genomes, as well as incongruences between species trees obtained using different methods or data partitions. 
Researchers are often left trying to make sense of such incongruences. 
Interpretive ways of measuring and visualizing discordance are needed, both among alternative species trees and gene trees, especially for specific focal branches of a tree. Here, we introduce DiscoVista, a publicly available tool that creates a suite of simple but interpretable visualizations. DiscoVista helps quantify the amount of discordance and some of its potential causes.
\end{abstract}
\section{Introduction}
The age of phylogenomics, once hoped to be the end of incongruence in phylogenetic analyses~\citep{Rokas2003}, has turned out to be ripe
with incongruence~\citep{Jeffroy2006} and methodological difficulties. 
The long-understood theoretical concerns about gene tree incongruence due to incomplete lineage sorting (ILS)~\citep{Maddison1997} have been implicated in many studies~\citep[e.g.,][]{avian,1kp-pilot,Suh2015}.
While methods
that seek to address gene tree incongruence have been developed,
no consensus has emerged as to the choice of the best methodology~\cite[see][for examples]{Edwards2016,Springer2015}.
Nevertheless, phylogenomics studies have to at least consider
the possibility of gene tree incongruence and its impacts, 
a feat made difficult by the noisy
estimates of gene trees~\citep{Springer2015,Patel2013,mrl-sysbio,localpp}. 
Moreover, in some cases, the incongruence itself may be of interest~\citep{Hahn2016}. 
Even ignoring gene tree discordance, 
choices of models to apply to the sequences, delineation of data partitions, alignment techniques, or simply
software packages used to analyze the data have all proved consequential~\cite[e.g.,][]{Jeffroy2006,avian,1kp-pilot,Romiguier2013,Philippe2011,Zwickl2014}.

These difficulties have compelled some researchers 
to use several alternative models and methods
and then test the sensitivity of results 
to such choices. Occasionally, analysts choose to also perturb the set of species included, and they often run analyses on different partitions of the data. 
The analyst hopes for congruence between various analyses that would indicate rubustness of the results to assumptions, but often observes differences. Ideally, the results of \textit{all} analyses should be published,
to convey the existence of incongruence in results to the reader.

As long as incongruence remains an important force in phylogenetics, we need interpretable ways to measure and visualize the discordance between species tree 
estimates resulting from different analytical method and assumptions, and also between gene trees and a  summary species tree. 
Sophisticated tools have been developed to visualize discordance. For example, DensiTree~\citep{densitree} overlays trees on top of each other to create a phylogenetic cloud, and
SplitsTree~\citep{splitstree} conveys incongruence by
producing a network while keeping some of the tree-like structure.
These tools create creative and striking
indicators of discordance. Yet 
it is often hard to interpret the meaning of such figures in measurable ways. 
We believe that in addition to these methods, phylogenomics will benefit from simple, interpretable, and easy-to-perform visualizations 
that help systematists to identify discordance and its potential causes.

In this paper, we introduce DiscoVista,
a tool that creates a series of simple yet powerful visualizations of discordance. 
DiscoVista is a command-line tool and relies on several other packages, including Dendropy~\citep{dendropy}, ape~\citep{ape}, newick utilities~\citep{newickutility}, the ggplot package~\citep{ggplot2}, 
and ASTRAL~\citep{localpp}.
The code, a Docker~\citep{Boettiger2015AnResearch} virtual image (for easy installation), 
and examples are available online at~\url{https://github.com/esayyari/DiscoVista}.

DiscoVista  can generate several 
visualizations (Table~\ref{tab:discovista}) that
summarize gene tree discordance and discordance among species trees, show taxon occupancy, and show sequence statistics such as GC content. 
DiscoVista strives for interpretability. 
In many analyses, not all branches in a phylogenetic tree are equally important because questions of interest typically 
concern several hypotheses surrounding
the relationships between focal groups.
Visualizing discordance with respect to only these focal relationships simplifies interpretation. 
Assessing hypotheses concerning these larger subsets of the species helps in answering the downstream biological questions of interest. 
Thus, DiscoVista allows researchers to define focal groups of taxa
and evaluate discordance relevant to those groups.

\begin{table}[tbp]
\centering
\caption{Description and examples for different DiscoVista analyses}
   \label{tab:discovista}
   \begin{tabular}{l|l|l}
     \hline
\textbf{Analysis Name} & \textbf{Shows $\ldots$} &  \textbf{Examples} \\ \hline
\multirow{2}{*}{Species tree compatibility} & compatibility of focal groups of species  & \multirow{2}{*}{Figs.~\ref{fig:species}a,~\ref{fig:FNA2AA.shade}, and~\ref{fig:FNA2AA.block}} \\ &in species tree &\\ \hline
\multirow{2}{*}{Occupancy analysis} & taxon occupancy for individuals or focal   & \multirow{2}{*}{Figs.~\ref{fig:species}bc and~\ref{fig:1kp-1}} \\
& groups of species in genes  & \\  \hline
\multirow{2}{*}{Gene tree compatibility }&  compatibility of focal groups of species&
\multirow{2}{*}{Figs.~\ref{fig:genes}a,~\ref{fig:geneNumber}, and~\ref{fig:geneFreq} }\\ 
& in gene trees & \\ \hline
\multirow{2}{*}{Branch quartet frequencies} &  quartet frequencies around important   & \multirow{2}{*}{Figs.~\ref{fig:genes}b and~\ref{fig:1kp-relfreq}} \\ 
 & branches of the species tree &  \\ \hline

GC content & GC content of each codon position & Figs.~\ref{fig:1kp-2}\\ \hline 
\end{tabular}
\end{table}

\section{Results}

We apply DiscoVista on three datasets to demonstrate its output visualizations.
The exact commands for generating example figures are given in the supplementary materials.

\paragraph{Datasets} 

\begin{figure*}[!htbp]
\centering
\includegraphics[width=0.83\textwidth]{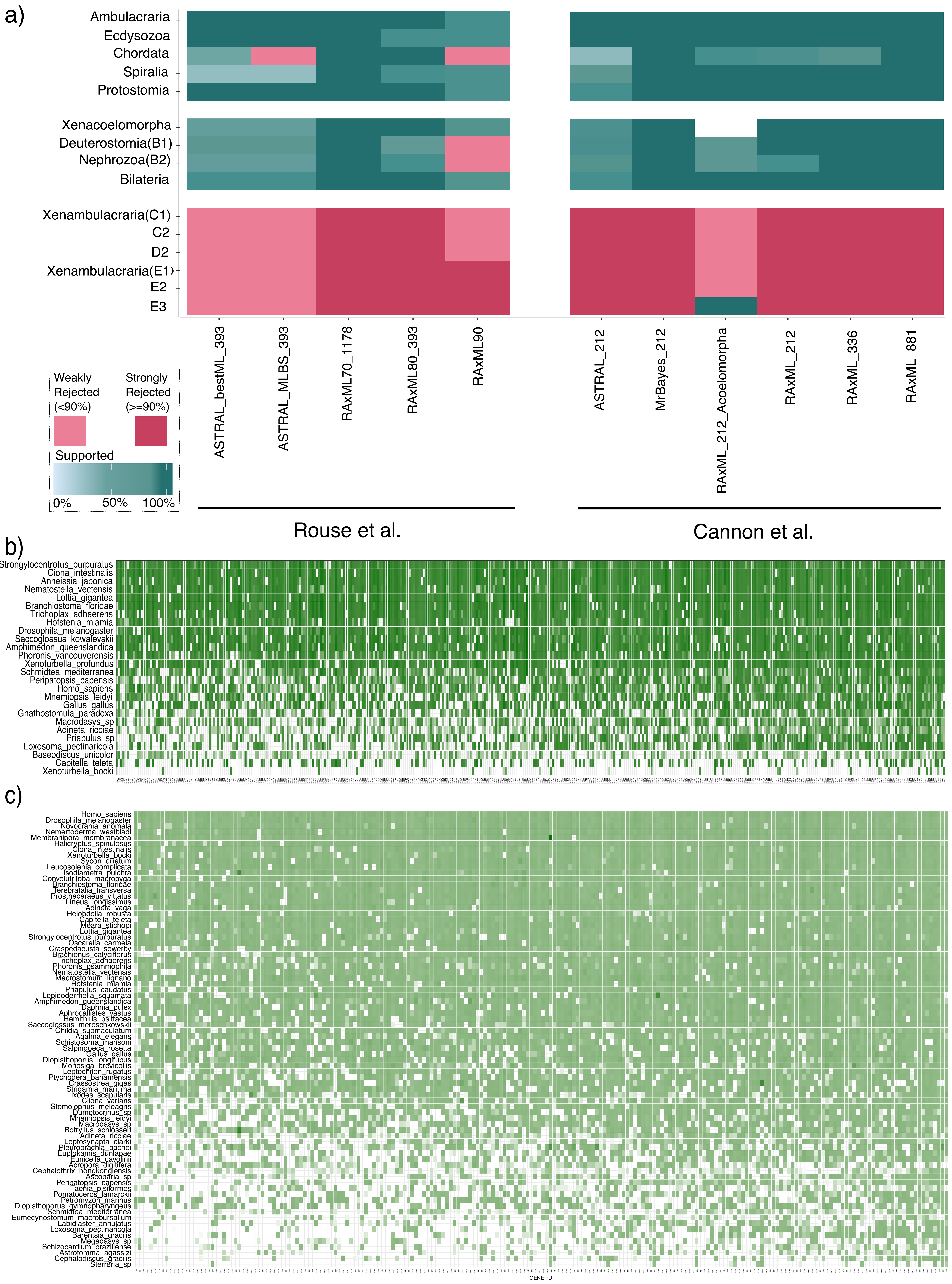}
\begin{center}
\caption{{\bf Species tree discordance and occupancy.} a) DiscoVista Specie tree analysis. Rows correspond to focal splits, and columns correspond to alternative species trees reported in two papers~\citep{Rouse2016,Cannon2016}. The spectrum of blue-green indicates support values for splits compatible with a tree. 
Note that support values are of different types (Bayesian posterior, concatenation bootstrap support, and multi-locus bootstrap support) and thus may not be directly comparable.  Weakly rejected splits correspond to splits that are not present in the tree, but are compatible if low support branches (below 90\%) are contracted. In~\citet{Rouse2016}, RAxML70\_1178, RAxML80\_393, and RAxML90 are results of concatenation analyses on genes with average occupancy of 70\%, 80\%, and 90\% respectively. For~\citet{Cannon2016},  RAxML\_212, RAxML\_212\_Acoelomorpha, RAxML\_336, and RAxML\_881 are concatenation analyses on 212 orthologus genes of 78 species, 212 orthologus genes after removing Acoelomorpha, 336 orthologus genes on 56 selected species, and 881 orthologus genes on 77 species, respectively. The ASTRAL\_212 is the species tree estimated using ASTRAL and 212 orthologus genes of 78 species. b,c) The occupancy map of 393 (b) and 212 (c) gene alignments of~\citet{Rouse2016} (b) and~\citet{Cannon2016} (c) with average occupancy of 80\% in both datasets. The spectrum of green color shows the gene length, and the white color indicates missing data.}
\label{fig:species}
\end{center} 
\end{figure*}

\begin{figure*}[!htb]
\centering
\includegraphics[width=0.75\textwidth]{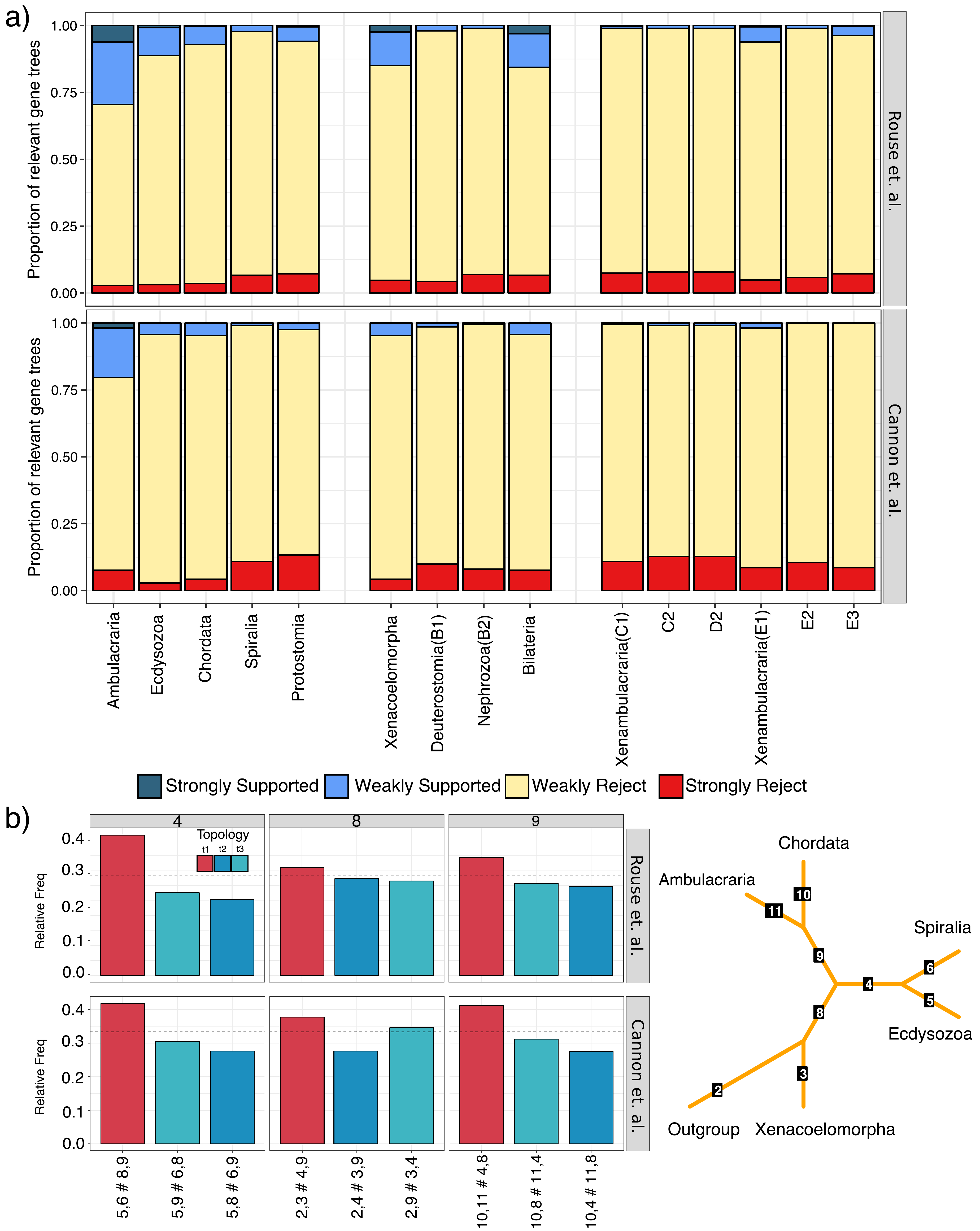}
\begin{center}
\caption{{\bf Gene tree discordance figures}. 
a) Gene tree compatibility. This figure shows the portion of RAxML genes for which focal splits (x-axis) are highly (weakly) supported or rejected. Weakly rejected splits are those that are not in the tree but are compatible if low support branches (below 75\%) are contracted. 
b) Frequency of three topologies around focal internal branches of ASTRAL species trees in both datasets. Main topology is shown in red, and the other two alternative topologies are shown in blue. The dotted line indicates the 1/3 threshold. The title of each subfigure indicates the label of the corresponding branch on the tree on the right (also generated by DiscoVista). Each internal branch has four neighboring branches which could be used to represent quartet topologies. On the x-axis the exact definition of each quartet topology is shown using the neighboring branch labels separated by ``{\#}".}
\label{fig:genes}
\end{center} 
\end{figure*}

The position of Xenacoelomorpha among deep branches of the Metazoan phylogeny has proved challenging to resolve, 
with two prevailing hypotheses.
One hypothesis puts Xenacoelomorpha as sister to all other Bilateria, while the other hypothesis puts them  inside Deuterostomia, implying a dramatic loss of complexity.
 Intriguingly, these marine worms are bilaterally symmetrical but lack several other features compared to most other bilaterians.  
Two independent and simultaneous studies by~\citet{Rouse2016} and~\citet{Cannon2016} have focused on the position of Xenacoelomorpha in the tree of life. These two studies used different (but overlapping) set of species and each analysis used several reconstruction methods. 
The two papers come to the same conclusion, putting Xenacoelomorpha as the sister to all other Bilateria.  
The final results of~\citet{Cannon2016} is based on 78 species and  212 orthologous genes with average per taxon occupancy of 80\%. 
Their paper includes alternative analyses based on several subsets of taxa and reconstruction methods. 
The dataset by~\citet{Rouse2016} includes 26 species and 1178 genes (including four Xenoturbella species) with average gene occupancy of 70\%. They also report alternative trees using concatenation and ASTRAL (run on 393 genes with 80\% occupancy). 
Although these two datasets used different set of taxa, 
by focusing on focal splits,
DiscoVista can generate visual comparisons of results across both datasets (Figs.~\ref{fig:species} and~\ref{fig:genes}). 

As a second example, we show DiscoVista results on a phylotranscriptomic dataset of 103 plants~\citep{1kp-pilot} in the supplementary material (Figs. \ref{fig:FNA2AA.shade} -- \ref{fig:1kp-2}).
This dataset comes with both DNA and AA sequences, allowing us to show additional figures
that could not be built for the Xenacoelomorpha datasets. 

\paragraph{Split definitions} 
A central input to DiscoVista is a {\em split definitions} file where the user can combine taxa into groups of interest and give names to the groups (see supplementary material for details).
Each split is a bipartition of the taxa into two groups and corresponds to
an edge in an unrooted tree. The user can specify one side of a split (which would be a clade if the side that doesn't include the root is given).
With careful definition of splits, alternative hypotheses of interest could be
specified. 
For our two empirical datasets, we are considering focal groups of species from original publications ~\citep{Rouse2016, Cannon2016}, as shown partially in Table~\ref{tab:Table1} (see Figs.~\ref{fig:clade-def} and~\ref{fig:clade-def2} for full definitions).

\begin{table}[tbp]
\centering
\caption{An example {\em split definition} file for the two Xenoturbella datasets; several lines on top define base splits and are left blank here, but see the full files in Figures~\ref{fig:clade-def} and~\ref{fig:clade-def2}.}
   \label{tab:Table1}
   \begin{small}
   \begin{tabular}{|l|l|}
     \hline
Cluster Name & Definition  \\ 
     \hline 
Ambulacraria & ...\\ \hline
Ecdysozoa & ...\\ \hline
Acoelomorpha & ...\\ \hline
Spiralia & ...\\ \hline
Xenacoelomorpha &  ...\\ \hline
Chordata & ...\\ \hline
Protostomia & Spiralia+Ecdysozoa\\ \hline
Deuterostomia & Ambulacraria+Chordata \\ \hline
Nephrozoa & Protostomia+Deuterostomia \\ \hline
Bilateria	 & Nephrozoa+Xenacoelomorpha \\ \hline
Xenambulacraria(C1)  &  Ambulacraria+Xenacoelomorpha\\ \hline
C2 & Chordata+Xenambulacraria(C1)\\ \hline
Bilateria(C3) & C2+Protostomia \\ \hline
D2 & Xenacoelomorpha+Deuterostomia \\  \hline
Bilateria(D3) & D2+Protostomia\\ \hline
Xenambulacraria(E1) & Xenoturbella+Ambulacraria \\ \hline
E2  &  Chordata+Xenambulacraria(E1) \\ \hline
E3  &  Protostomia+E2 \\ \hline
\end{tabular}
   \end{small}
\end{table}

\paragraph{i) Species tree compatibility}
This visualization shows whether focal splits are supported, weakly rejected, or strongly rejected by different analyses. 
The inputs are a set of species trees in newick format,
a support threshold, and the splits definition file; the output is a heat map. 
For each focal split and for each species tree, if the split is  compatible with the tree, the corresponding cell is in shades of blue/green, and the spectrum of blue-green color indicates the branch support (any measure of support, including the bootstrap support, Bayesian posterior probability, or localPP~\citep{localpp} values could be used.) 
A split is considered weakly rejected if it is incompatible with the fully resolved species tree, but is {\em compatible} with the tree when branches with support values below the input threshold (e.g., 75\% BS) get contracted.
Compatibility is 
a concept that we find useful for measuring discordance in an interpretable way. 
A split is compatible with an unrooted tree if it is compatible with all splits of that tree and can therefore be added
to that tree; compatibility can be easily and efficiently checked~\citep{Warnow-treecompat}. 
Let two splits be $A|B$ 
and $C|D$; then, the two 
splits are compatible if 
and only if one of the four pairwise intersections $A \cap C$, $A \cap D$,
$B\cap C$, or $B \cap D$ is empty.

On the empirical datasets, the species tree compatibility figures (Fig.~\ref{fig:species}a) show several patterns. 
The two datasets produce largely congruent results, with only minor differences.  
In the~\citet{Rouse2016} dataset, aggressive filtering of genes based on their occupancy negatively impacts concatenation trees in terms of BS support and the recovery of some clades (e.g., Chordata). 
Consistent with literature~\citep{mrl-sysbio}, ASTRAL run
on maximum likelihood gene trees (bestML) seems more accurate than ASTRAL
run on the bootstrap replicates (MLBS) gene trees (impacting the recovery of Chordata). 

\paragraph{ii) Occupancy analysis}
Missing data is common in phylogenomics, and common filtering practices can further increase the amount of missing data with potential impact on downstream analyses~\cite[e.g.,][]{Hosner2016,Mai2017}. DiscoVista can visualize the taxon occupancy for individual species or for groups of taxa; taxon occupancy is 
defined as the fraction of gene alignments that have at least one of the species from a group. 
The inputs to this analysis are a set of sequence alignment files (FASTA format), and a species annotation file; 
the output is a line plot of the occupancy per species or per group. 
DiscoVista can also visualize taxon occupancy as a heatmap that shows the presence/absence of species in gene alignments, and for those present, uses color shades to indicate their non-gapped sequence length compared
to the other sequences of the same gene. 

On the empirical \citet{Rouse2016} dataset, the occupancy plots immediately reveal how {\em Xenoturbella bocki} has low occupancy (Fig.~\ref{fig:species}b), appearing only in a handful of genes. 
However, the negative impacts of this low occupancy may be ameliorated by the high occupancy of the other two Xenacoelomorpha species, \textit{Xenoturbella profoundus} and \textit{Hofstenia miamia}. 
Patterns of occupancy do not vary widely on this dataset.
On the Cannon dataset, in contrast, some genes have lower levels of occupancy than others (from left to right in Fig.~\ref{fig:species}c). 
Just like the Rouse dataset, several species have very low occupancy (e.g., {\em Astrotomma agassizi, Cephalodiscus gracilis, Sterreria sp.}). However, {\em Xenoturbella bocki}, which is the only Xenacoelomorpha species in this dataset, has high occupancy and appears in most of the genes. 

\paragraph{iii) Gene tree compatibility}
This visualization simply depicts the portion of gene trees supporting or rejecting each focal split. 
The inputs are one or several collections of gene trees, the split definition file, and a support threshold. 
For splits that are compatible with gene trees, branches with bootstrap support values above (below) a threshold are considered as highly (weakly) supported. 
Splits that are not compatible with the original tree but are compatible with the tree if branches with low support values get contracted are considered as weakly rejected splits, and those that are not compatible even after contracting low support branches are considered strongly rejected. 
By default, gene trees that miss one side of the split completely are removed from the analysis.
The figure can also be created such that genes that completely miss one side of the split (or completely miss a user-defined subset of a side of the split) are marked as ``missing'' (Fig.~\ref{fig:geneNumber}). 

On the empirical datasets, these figures reveal high levels of gene tree discordance (Fig.~\ref{fig:genes}a).
Although ASTRAL species trees on maximum likelihood gene trees are mostly congruent in both datasets and with concatenation results, gene trees show high amounts of discordance, and none of the major splits are recovered in most gene trees (Fig.~\ref{fig:genes}a). 
However, for all of the focal splits, the majority of the gene trees are compatible with the species tree after contracting low support branches (below 75\%). 
Interestingly, maximum likelihood gene trees in~\citet{Rouse2016} show somewhat less discordance with major species tree splits compared to those in \citet{Cannon2016}.

\paragraph{iv) Branch quartet frequencies} Every internal branch of a species tree divides the tree into four parts, and thus, at least one quartet tree (often many) can be mapped uniquely onto that branch. For a given branch, assuming (a) the branch is correct (b) the only source of discordance between gene trees and species tree is ILS, and (c) there is no gene tree estimation error,  the multi-species coalescent (MSC) model has specific expectations about these quartets~\citep{Allman}. 
The probability of a gene tree quartet tree matching the species tree topology ($p>\frac{1}{3}$) is higher than probability of matching the two alternatives, and the two alternatives have equal probabilities ($q=\frac{1-p}{2}<\frac{1}{3}$).
Visualizing the empirical frequency of quartets around each branch can serve two purposes: it gives an interpretable measure of discordance specific to that branch~\citep{localpp}, and one can check whether the assumptions of ILS are met. 
If the discordance is purely due to ILS, then one would expect the second and the third hypotheses to have similar frequencies (close to $q$).
Lack of this pattern can have many causes, including
hybridization~\citep{Solis-Lemus2016}.
Finally, note that if the length of a species tree branch in coalescent units~\citep{Allman} is $d$, then for quartets around it, $p=1-\frac{2}{3}e^{-d}$, and thus, the quartet frequencies also convey information about the branch length. 
In the limit, for $d=0$ (e.g., a polytomy) one would expect all three frequencies to be close to $\frac{1}{3}$ and $p=q=\frac{1}{3}$ can be tested as a null hypothesis~\citep{polytomy}.

In this visualization, for every focal split, a bar graph shows the quartet frequencies around that branch. 
Moreover, a cartoon tree is generated where leaves are the large groups
of taxa collapsed into single leaves and branches are labeled
consistently with the bar graph. 
Inputs to this analysis are a set of gene trees, a species tree, an annotation file that maps each species to a named group and thus defines leaves of the cartoon tree
and by extension, the focal splits.

The portion of quartets around three focal branches of the ASTRAL species trees in our empirical studies show why some branches have been controversial (Fig.~\ref{fig:genes}b). 
For Nephrozoa (branch labeled as ``8'') and Deuterostomia (labeled as ``9''), the relative frequency of main topologies are extremely close to 1/3 in both studies. 
This high level of gene tree discordance around these two branches is likely caused by a combination of high true discordance and also gene tree estimation error; irrespective of which is more prevalent, 
the high discordance reveals a cause of difficulties faced in resolving these relationships. 
Interestingly, the portion of quartets for alternative topologies in~\citet{Rouse2016} follow the expectations of the MSC theory quite well (i.e., second and third topologies have very close frequencies); the same cannot necessarily be said of the~\citet{Cannon2016} gene trees.

\paragraph{v) GC content} 
Commonly used models of sequence evolution are
stationary and  assume that all species have identical base composition. 
This assumption is often violated
in gene coding sequences, especially
in the third codon position~\citep{Jeffroy2006}.
While non-stationary models~\citep{nhPhyml} and tests
of divergence from stationarity~\citep{Ababneh2006} 
exist,
simply visualizing GC content of different codon position for different extant taxa can help in judging 
non-stationarity and in deciding whether
a GTR analysis is appropriate for some or all of the data~\citep{avian,1kp-pilot}.
DiscoVista generates box plots (distributions) or dot plots (averages)  of GC content of each codon position, as well as all three codon positions, for each species. The input to this analysis is a set of gene coding sequences in FASTA format.

The two Xenacoelomorpha datasets did not release DNA sequences and therefore, we could not compute their GC content. 
Nevertheless, examples of the GC content figures could be seen for the plant dataset  (Fig.~\ref{fig:1kp-2}). These figures immediately show that assumptions of equal base frequencies are violated. 
They also show that the variations in the GC content are mostly concentrated in the third codon position. The GC content levels for 
the second codon position, and to a lesser degree the first codon position, do not vary much across species. 
These results favor the removal of the third codon position when building gene trees or in concatenation analyses. 


To summarize, DiscoVista provides a useful tool for visualizing several patterns of discordance,
missing data, and GC variations in phylogenomic datasets.

\bibliographystyle{natbib}
\bibliography{smmendeley,discovista}

\clearpage

\begin{center}\Large
Supplementary Material for ``DiscoVista: interpretable
visualizations of gene tree discordance''
\par\end{center}

\begin{center}\small
Erfan Sayyari\textsuperscript{1},
James B. Whitfield\textsuperscript{2},
Siavash Mirarab\textsuperscript{1*}
\end{center}

\begin{center}\small
$^{1}$Department of Electrical and Computer Engineering,
University of California at San Diego, 9500 Gilman Drive, La Jolla, CA 92093. 

$^{2}$Department of Entomology, 320 Morrill Hall, University of Illinois, 505 S. Goodwin Avenue, Urbana, IL 61801
\end{center}

\setcounter{equation}{0}
\setcounter{figure}{0}
\setcounter{table}{0}
\setcounter{page}{1}
\makeatletter
\renewcommand{\theequation}{S\arabic{equation}}
\renewcommand{\thefigure}{S\arabic{figure}}
\renewcommand{\bibnumfmt}[1]{[S#1]}
\renewcommand{\citenumfont}[1]{S#1}

\clearpage

This document provides supplementary 
tables and figures used in the main paper.

\section*{Supplementary Figures}

\begin{figure*}[!htbp]
\begin{center}
\includegraphics[width=0.95\textwidth]{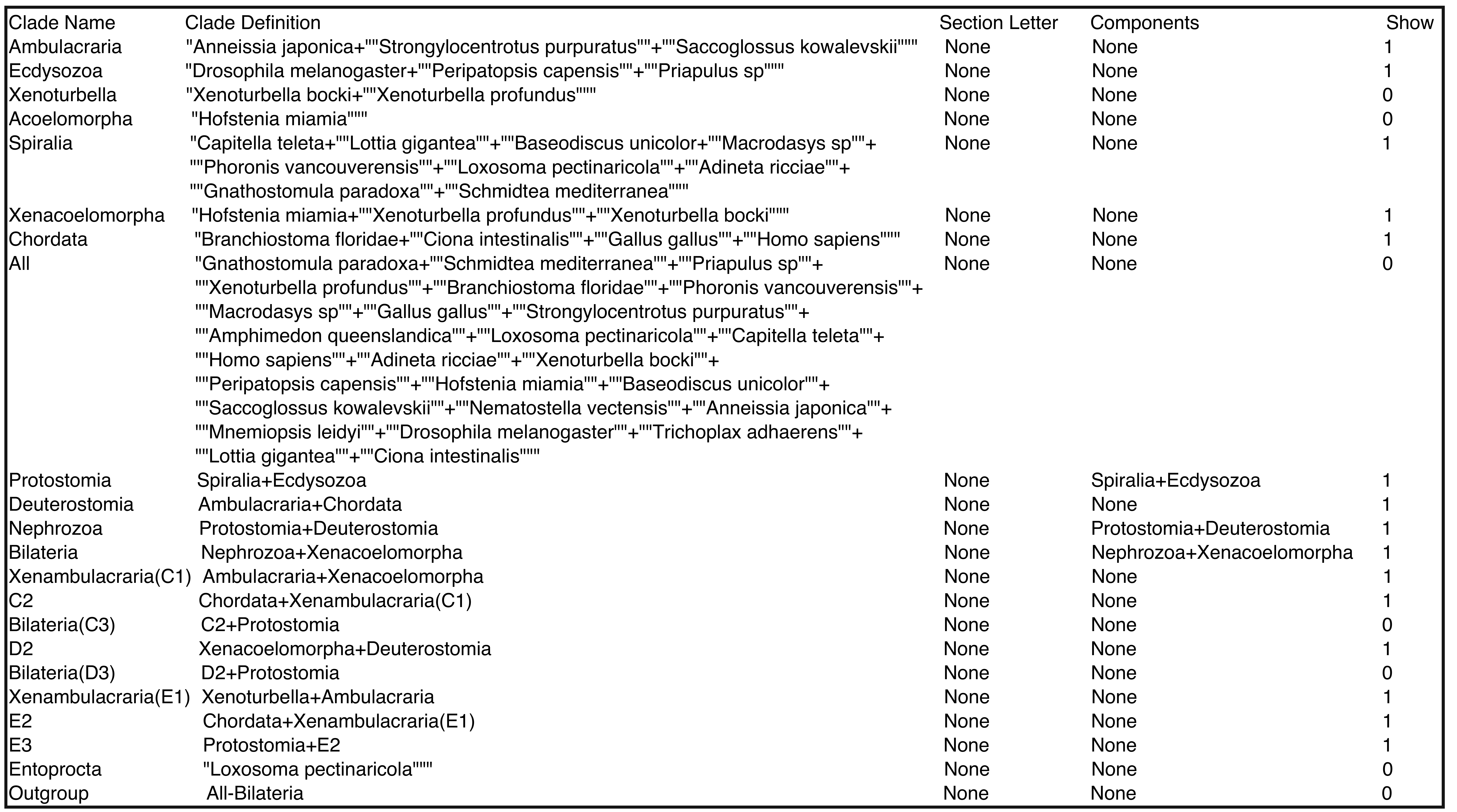}
\end{center}
\caption{Complete split definitions for the Xenoturbella dataset of~\citet{Rouse2016}.}\label{fig:clade-def}
\end{figure*}

\begin{figure*}[!htbp]
\begin{center}
\includegraphics[width=0.95\textwidth]{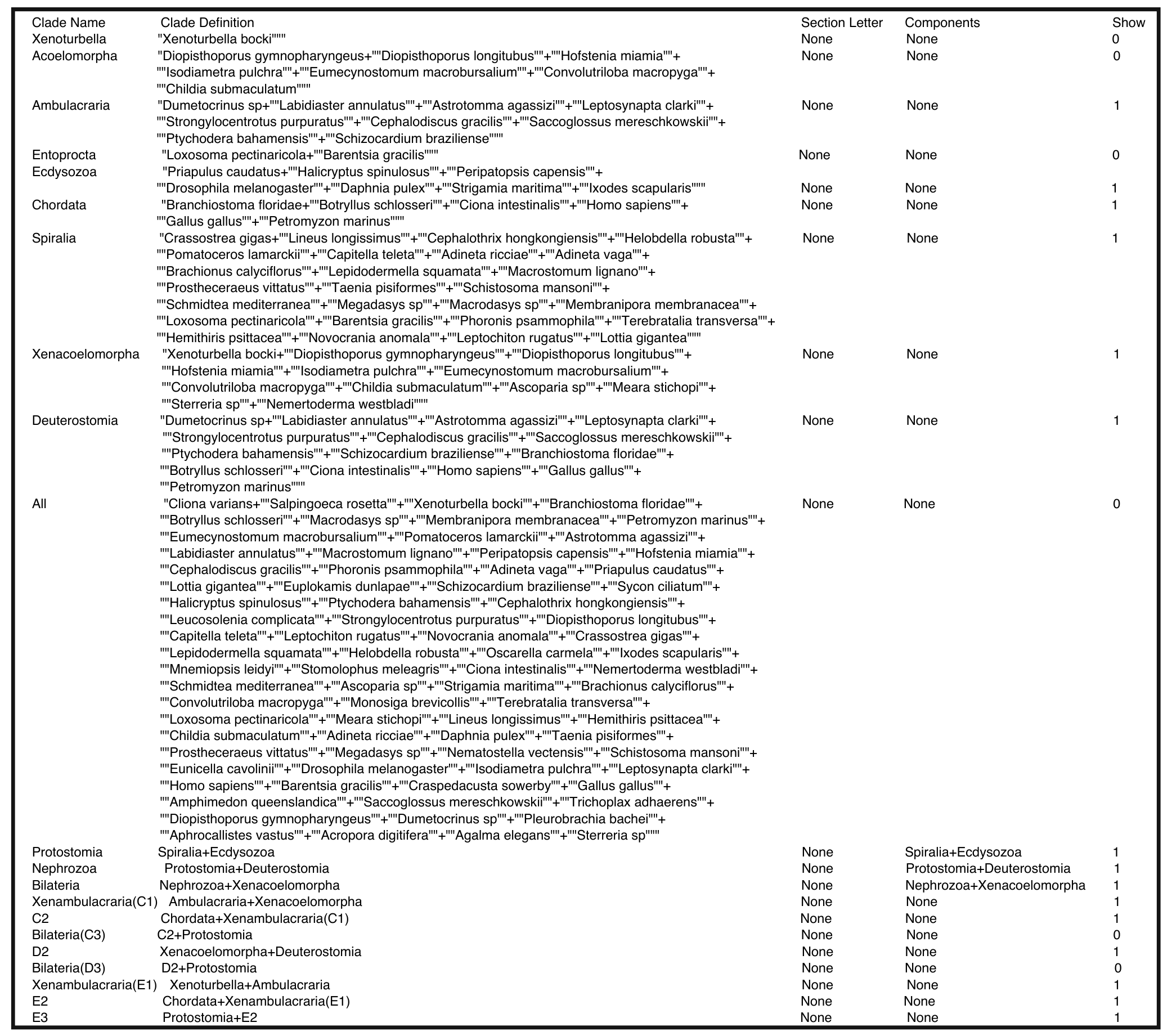}
\end{center}
\caption{Complete split definitions for the Xenoturbella dataset of~\citet{Cannon2016}.}\label{fig:clade-def2}
\end{figure*}

\newpage 

\begin{figure*}[!htbp]
\begin{center}
\includegraphics[width=1\textwidth]{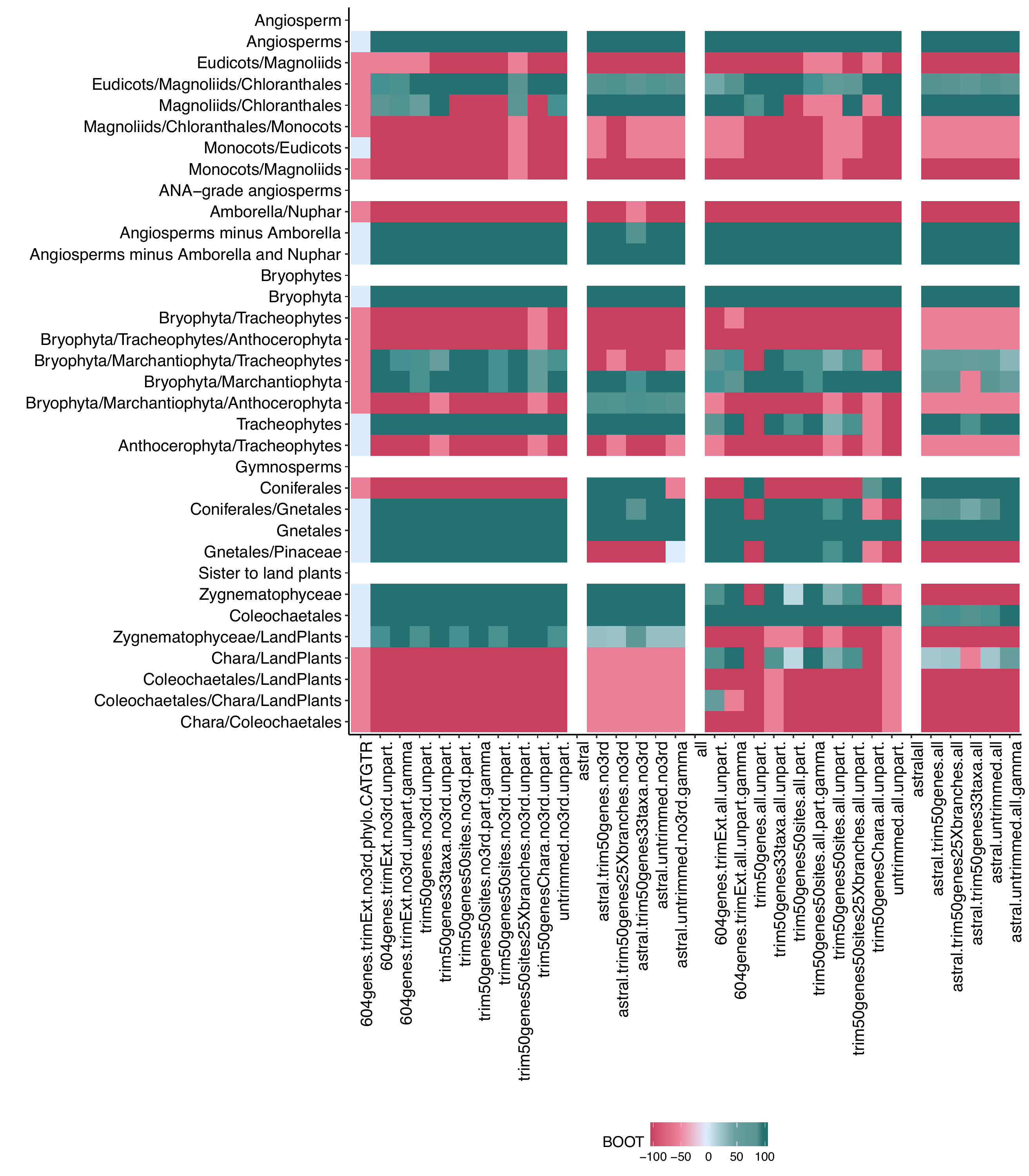}
\end{center}
\caption{
DiscoVista Specie tree analysis on 1kp dataset: Rows correspond to major orders and clades, and columns correspond to the results of different methods reported in two different closely related datasets. The spectrum of blue-green indicates amount of MLBS values for monophyletic clades.  Weakly rejected clades correspond to clades that are not present in the tree, but are compatible if low support branches (below 90\%) are contracted. }\label{fig:FNA2AA.shade}
\end{figure*}

\begin{figure*}[!htbp]
\begin{center}
\includegraphics[width=1\textwidth]{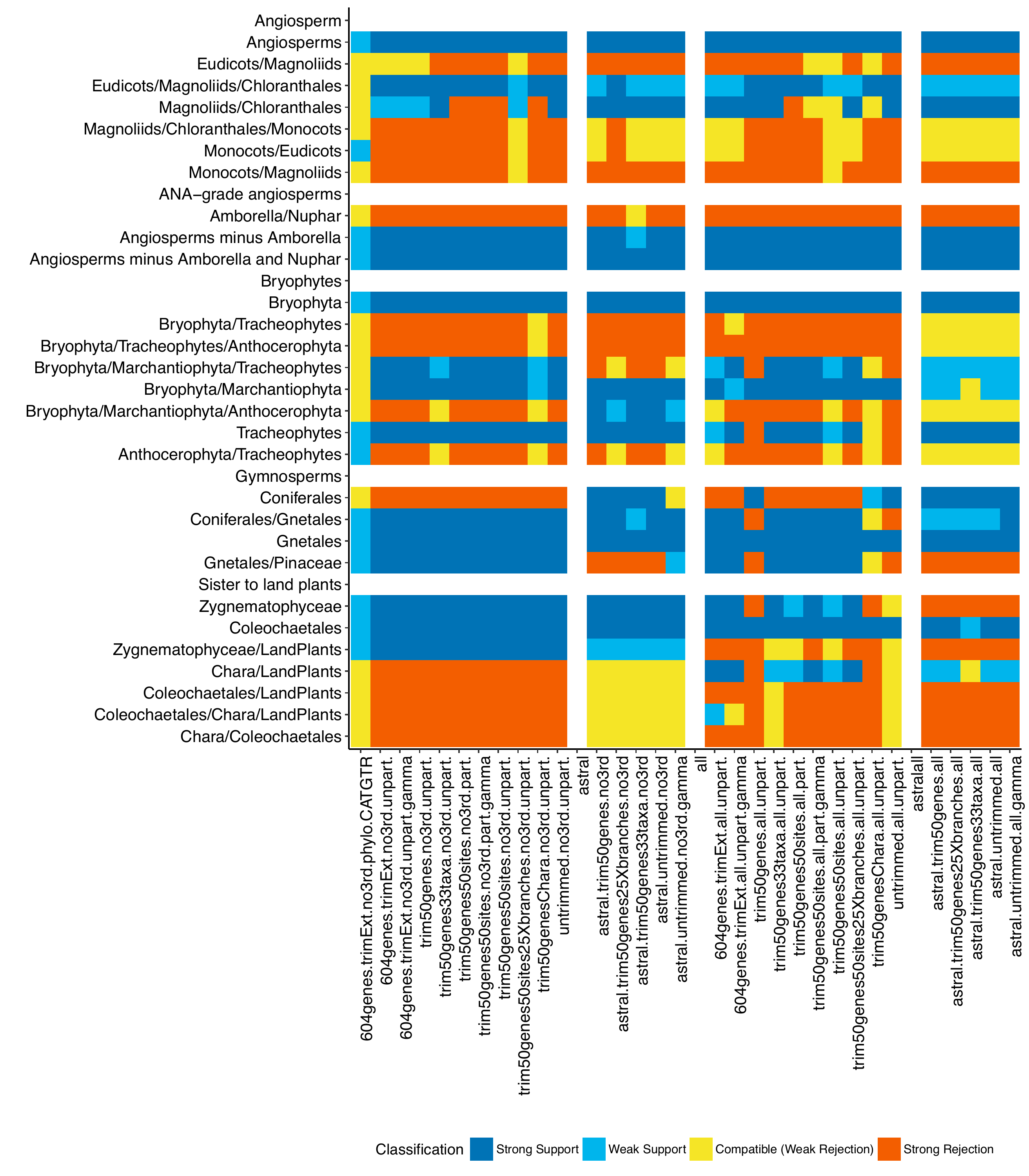}
\end{center}
\caption{DiscoVista Specie tree analysis on 1kp dataset: Rows correspond to major orders and clades, and columns correspond to the results of different methods reported in two different closely related datasets.  Weakly rejected clades correspond to clades that are not present in the tree, but are compatible if low support branches (below 90\%) are contracted.}\label{fig:FNA2AA.block}
\end{figure*}

\begin{figure*}[!htbp]
\includegraphics[width=0.95\textwidth]{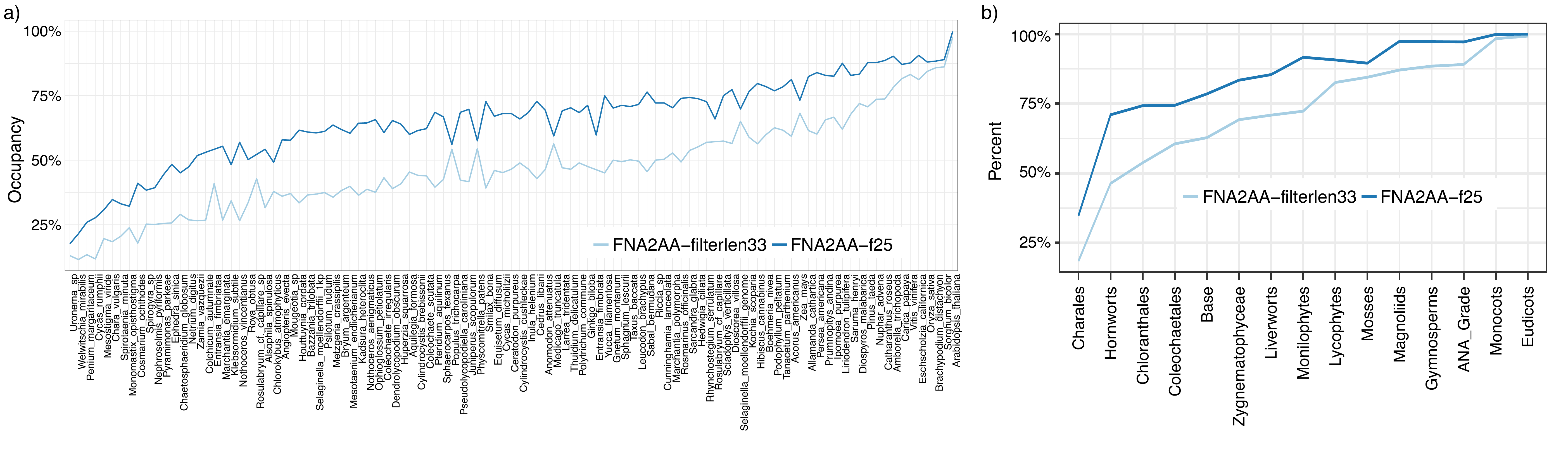}
\begin{center}
\caption{{DiscoVista analyses on 1kp dataset} a) occupancy analysis on the 1kp dataset over each individual species for two model conditions.  FNA2AA-f25: amino acid sequences back translated to DNA, and sequences on long branches (25X median branch length) removed; FNA2AA-filterlen33:  amino acids sequences back translated to DNA, and fragmentary sequences removed (66\% gaps or more). b) occupancy analysis of major clades in the 1kp dataset and the same model conditions as (a). 
}
\label{fig:1kp-1}
\end{center}
\end{figure*}

\begin{figure*}[!htbp]
\begin{center}
\includegraphics[width=0.75\textwidth]{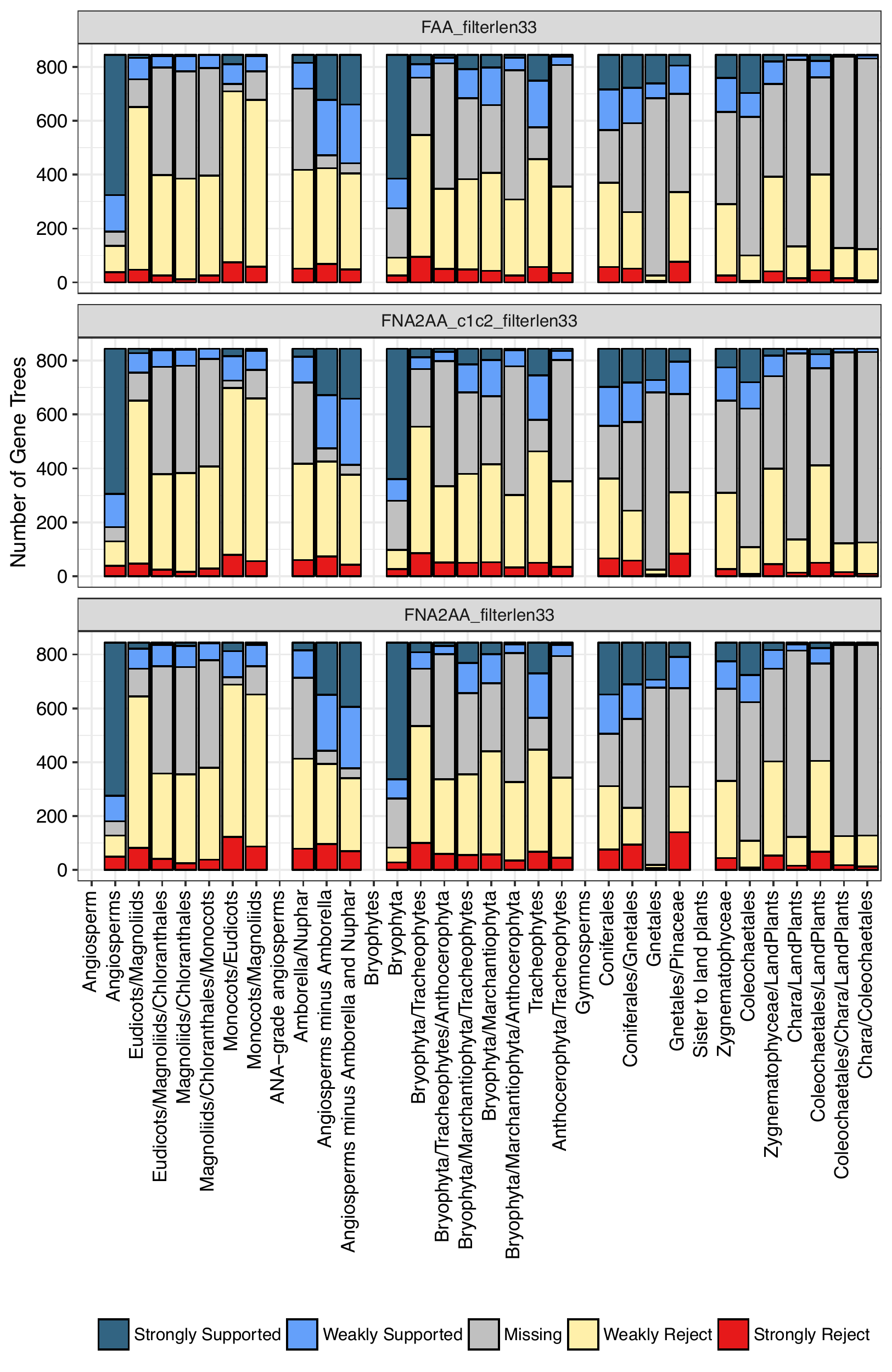}
\end{center}
\caption{Gene tree analysis on 1kp dataset: The portion of RAxML genes for which important clades (x-axis) are highly (weakly) supported or rejected for three model conditions of the 1kp dataset. FAA-filterlen33: gene trees on amino acids sequences, and fragmentary sequences removed (66\% gaps or more) FNA2AA-f25: amino acid sequences back translated to DNA, and sequences on long branches (25X median branch length)removed; FNA2AA-filterlen33: amino acid sequences back translated to DNA, and fragmentary sequences removed (66\% gaps or more). Weakly rejected clades are those that are not in the tree but are compatible if low support branches (below 75\%) are contracted}
\label{fig:geneNumber}
\end{figure*}

\begin{figure*}[!htbp]
\begin{center}
\includegraphics[width=0.8\textwidth]{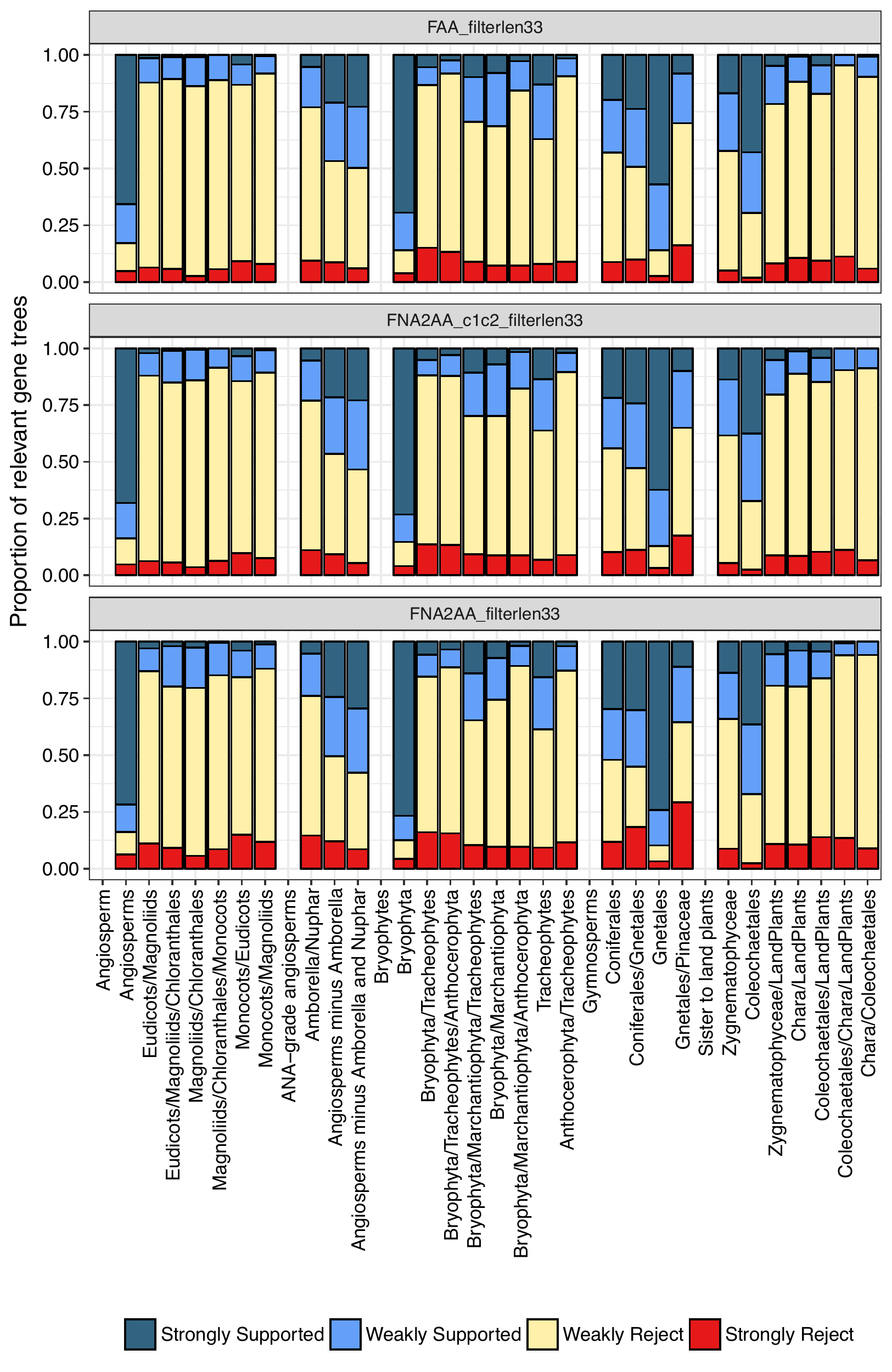}
\end{center}
\caption{Gene tree analysis on 1kp dataset: The number of RAxML genes for which important clades (x-axis) are highly (weakly) supported or rejected or are missing of three model conditions (same as~\ref{fig:geneNumber}) of the 1kp dataset. Weakly rejected clades are those that are not in the tree but are compatible if low support branches (below 75\%) are contracted.}
\label{fig:geneFreq}
\end{figure*}

\begin{figure*}[!h]
\includegraphics[width=0.8\textwidth]{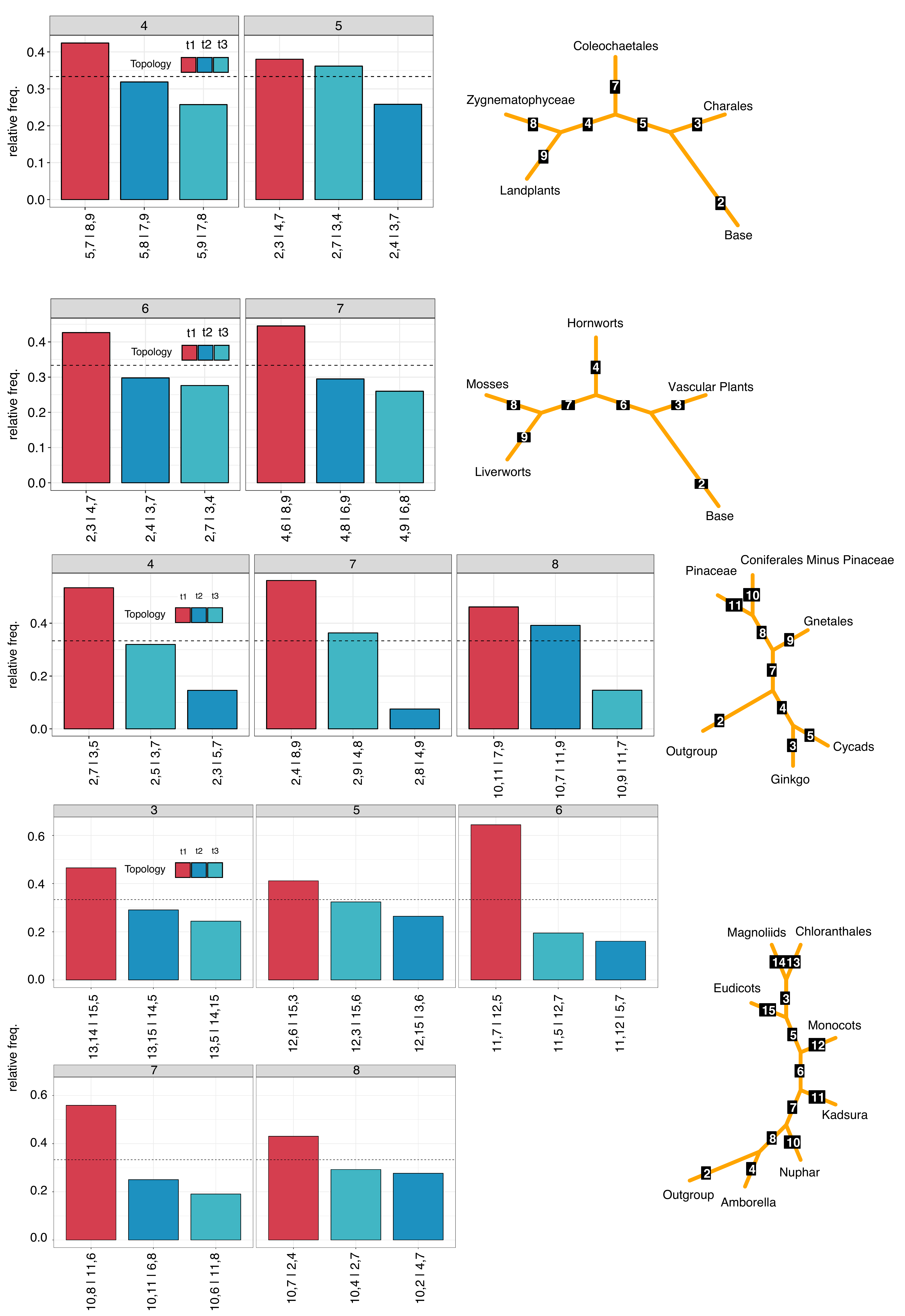}
\begin{center}
\caption{{Example DiscoVista visualizations of gene tree discordance on the 1kp dataset~\citep{1kp-pilot}.} Branch quartet frequencies graphs. Bars show the frequencies of observing the three quartet topologies around focal branches of the ASTRAL species tree among the main trimmed gene trees of the 1kp dataset. Each internal branch has four neighboring branches, which can be arranged in three ways. The frequency of the species tree topology among gene trees is shown in red, and the other two alternative topologies are shown in blue. The dotted lines indicate the 1/3 threshold. The title of each box indicates the label of the corresponding branch on the associated cartoon tree (also generated by DiscoVista). On the x-axis the exact definition of each quartet topology is shown using the neighboring branch labels separated by ``{\#}".} 
\label{fig:1kp-relfreq}
\end{center}
\end{figure*}

\begin{figure*}[!htbp]
\includegraphics[width=0.95\textwidth]{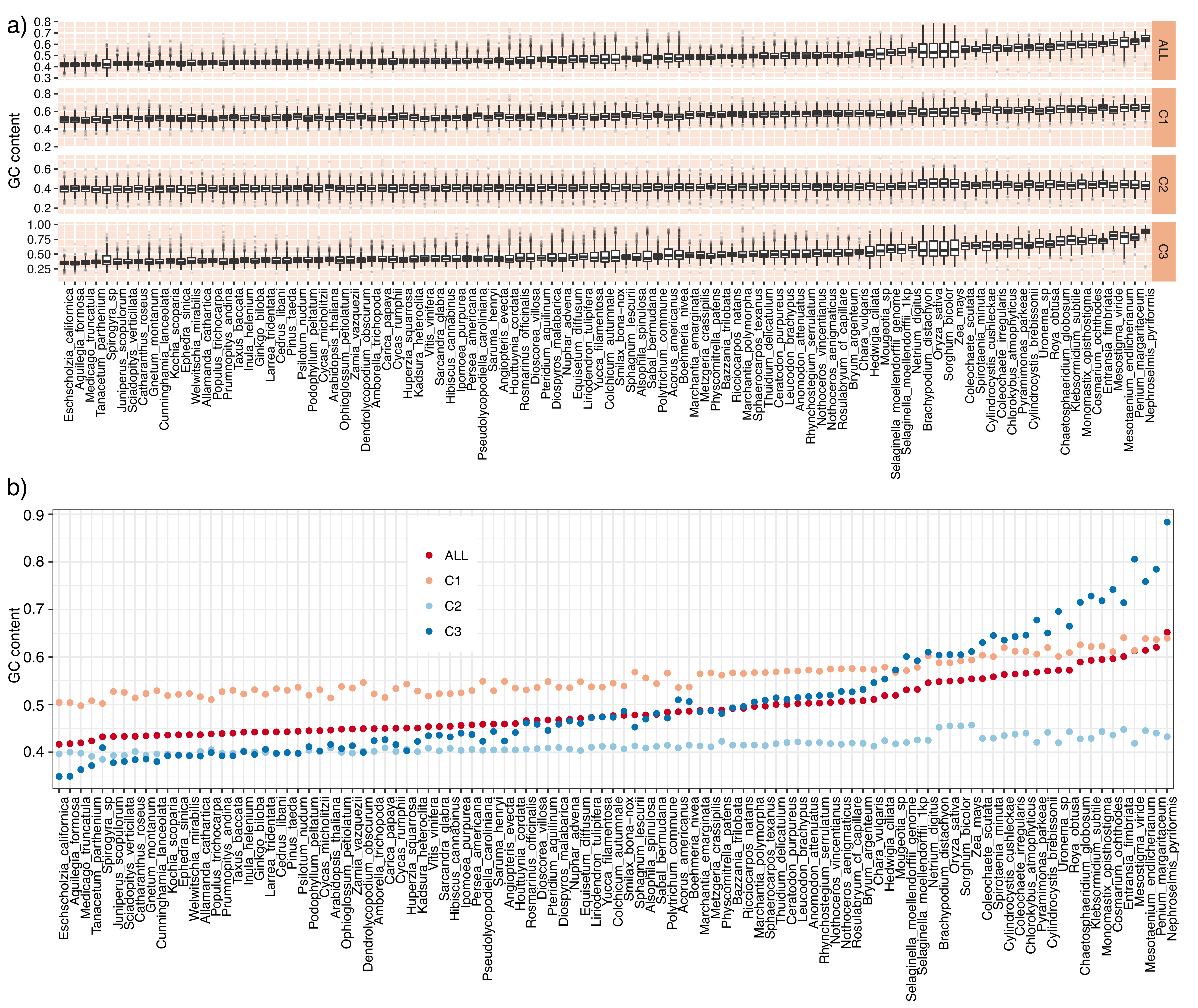}
\begin{center}
\caption{{DiscoVista analyses on 1kp dataset} a,b) GC content analysis of the 1kp dataset using boxplots and dot plot respectively for first, second, third, as well as all three codon positions. In dot-plot each dot shows the average GC content ratio for each species in all (red), first (pink), second (light blue), and third (dark blue) codon positions.
}
\label{fig:1kp-2}
\end{center}
\end{figure*}

Figures \ref{fig:FNA2AA.shade} to \ref{fig:1kp-2} are all for the plant dataset of \citet{1kp-pilot}. 

\clearpage

\section*{Structure of parameter files}\label{sec:struct}

\paragraph{splits definitions:} The splits definition file has 7 columns separated with tabs. In the first column split names are listed, in the second column the species or other splits that define this split are listed separated with ``+'' or ``-'' signs. ``+'' signs are used to add splits and species, and ``-'' signs to subtract species or splits. In the third column (you might leave this column blank) a name can be defined that is used to specify the part of the tree that the splits belongs to, e.g. 1-Base, or Base. Forth column defines splits components where in the absence of any of them splits is considered as missing. The fifth column indicates whether the split should be shown in the species tree (gene trees) analysis.  The sixth column is used to define components from the other side of the split, where in the absence of any of them the split is considered as missing. Note that if you leave this column blank, ``All'' minus this split will be considered as the other side component.  The last column is used to add any comments to this file. Also note that, a split with the name ``All'' should be defined which consists of all the species names in your analysis.

\paragraph{Annotation file:} In this file, there should be a row for each species. The first column of the annotation file specifies the name of species, and  the second column defines the split that this species belongs to. Columns are separated by tab.

\section*{Installation, Commands, and Usage}
In order to install DiscoVista from source please refer to ~\url{https://github.com/esayyari/DiscoVista}. DiscoVista relies on different R, python , some C packages, as well as ASTRAL. In order to make the installation easier we also provide a docker image available at ~\url{https://hub.docker.com/r/esayyari/discovista/}. In order to use this image, after installing docker, it is sufficient to download it using the following command:

\begin{verbatim}
docker pull esayyari/discovista 
\end{verbatim}

Now, you would run different analyses using the following command:

\begin{verbatim}
docker run -v <absolute path to data folder>:/data 
esayyari/discovista  discoVista.py [OPTIONS]
\end{verbatim}

\paragraph{Species discordance analysis:} For this analysis you need the path to the species trees folder, the clade definitions, the output folder, and a threshold. For more information regarding the structure of the species tree folder please refer to ~\url{https://github.com/esayyari/DiscoVista}. Then you would use the following command using docker:

\begin{verbatim}
docker run -v <absolute path to data folder>:/data esayyari/discovista  discoVista.py 
-m 0  -p <path to species trees folder> -t <threshold> -c <path to split definition> 
-o <output folder> -y <model condition ordering file> -w <splits ordering file>
\end{verbatim}
 
\paragraph{Discordance analysis on gene trees:} For this analysis you need the path to the gene trees folder, the clade definitions, the output folder, and a threshold. For more information regarding the structure of the gene trees folder please refer to ~\url{https://github.com/esayyari/DiscoVista}. Then you would use the following command with docker:
 
 \begin{verbatim}
 docker run -v <absolute path to data folder>:/data esayyari/discovista discoVista.py 
 -m 1 -p <path to gene trees folder> -t <threshold> -c <path to split definition> 
 -o <output folder> -w <splits ordering file>
\end{verbatim}

\paragraph{GC content analysis:} For this analysis you need the path to the coding alignments (in FASTA format) with the structure described in more details at ~\url{https://github.com/esayyari/DiscoVista}, and the output folder. You would use the following command in docker:
 \begin{verbatim}
 docker run -v <absolute path to data folder>:/data esayyari/discovista discoVista.py 
 -m 2 -p <path to alignments folder>  -o <output folder>
\end{verbatim}

\paragraph{occupancy analysis:} For this analysis you need the path to the alignments (in FASTA format) with the structure described in more details at ~\url{https://github.com/esayyari/DiscoVista}, and the output folder. Then you can use this command with the docker:
\begin{verbatim}
docker run -v <absolute path to data folder>:/data esayyari/discovista discoVista.py 
-p $path -m 3 -a <path to annotation file> -o <output folder> 
\end{verbatim}

\paragraph{Relative frequencey analysis:} For this analysis you need the path to the gene trees and species tree folder (structure described on the Github), annotation file, outgroup clade name, output folder. Using docker you would do this analysis with the following command:

\begin{verbatim}
docker run -v <absolute path to data folder>:/data esayyari/discovista discoVista.py 
-p $path -m 5 -a <path to annotation file> -o <output folder>
\end{verbatim}

If desired, the cartoon tree can be shown rooted, requiring  the name of the outgroup as input (e.g. Base or Outgroup). 

\begin{verbatim}
docker run -v <absolute path to data folder>:/data esayyari/discovista discoVista.py 
-p $path -m 5 -a <path to annotation file> -o <output folder> -g <outgroup clade name>
\end{verbatim}

\newcommand{\captionb}[2]{\caption[#1]{{\bf #1} #2}}


\clearpage

\end{document}